\documentclass[a4paper,11pt]{article}\usepackage{jcappub}\def\PRDstyle#1{}\def\JCAPstyle#1{#1}\let\Abstract\abstract
\usepackage[utf8]{inputenc}
\usepackage[T1]{fontenc}
\usepackage{cmap}

\def\imo{i}

\def\im#1{Im(#1)}
\def\Order#1{{\cal O}\left(#1\right)}
\newcommand{\ie}{{i.e.,}~}

\begin{document}

\title{How general is the strong cosmic censorship bound for quasinormal modes?}

\JCAPstyle{
\author[1]{R.~A.~Konoplya,}\emailAdd{roman.konoplya@gmail.com}
\author[2]{A.~Zhidenko}\emailAdd{olexandr.zhydenko@ufabc.edu.br}
\affiliation[1]{Research Centre for Theoretical Physics and Astrophysics, \\ Institute of Physics, Silesian University in Opava, \\ Bezručovo náměstí 13, CZ-74601 Opava, Czech Republic}
\affiliation[2]{Centro de Matemática, Computação e Cognição (CMCC), \\ Universidade Federal do ABC (UFABC), \\ Rua Abolição, CEP: 09210-180, Santo André, SP, Brazil}
\arxivnumber{2210.04314}
}

\PRDstyle{
\author{R. A. Konoplya}\email{roman.konoplya@gmail.com}
\affiliation{Research Centre for Theoretical Physics and Astrophysics, Institute of Physics, Silesian University in Opava, Bezručovo náměstí 13, CZ-74601 Opava, Czech Republic}
\author{A. Zhidenko}\email{olexandr.zhydenko@ufabc.edu.br}
\affiliation{Centro de Matemática, Computação e Cognição (CMCC), Universidade Federal do ABC (UFABC), \\ Rua Abolição, CEP: 09210-180, Santo André, SP, Brazil}
}

\Abstract{
Hod's proposal claims that the least damped quasinormal mode of a black hole must have the imaginary part smaller than half of the surface gravity at the event horizon.
The  Strong Cosmic Censorship in General Relativity implies that this bound must be even weaker: half of the surface gravity at the Cauchy horizon. The appealing question is whether these bounds are limited by the Einstein theory only?
Here we will present numerical evidence that once the black hole size is much smaller than then the radius of the cosmological horizon, both the Hod's proposal and the strong cosmic censorship bound  for quasinormal modes are satisfied for general spherically symmetric black holes in an arbitrary metric theory of gravity. The low-lying quasinormal frequencies  have the universal behavior in this regime and do not depend on the near-horizon geometry, but only on the asymptotic parameters: the value of the cosmological constant and black hole mass.
}

\maketitle

\section{Introduction}
The existence of the Cauchy horizon in a black-hole solution of the Einstein's or other alternative theory of gravity means that the future history of an observer crossing such a horizon cannot be known via integration of gravitational field equations starting from some initial data.
In other words, the gravitational theory looses determinism at the Cauchy horizon. For a charged black hole in the Einstein theory of gravity this problem is treated by the instability of the Cauchy horizon: perturbations outside the event horizon can be infinitely amplified inside the black hole by a blueshift mechanism, transforming the Cauchy horizon into the singularity where the usual Einstein-Maxwell equations are invalid and must be replaced with some improved theory. Mathematically, the Penrose’s Strong Cosmic Censorship conjecture (SCC) supports the latter observation stating that the maximal Cauchy development of generic compact or asymptotically flat initial data is locally inextendible as a regular Lorentzian manifold \cite{Penrose:1969pc}.

At the same time, viable model of a black hole must be stable against spacetime perturbations, so that the balance between the decay of perturbations outside the event horizon (expressed in the decay rate of the least damped \textit{quasinormal mode} \cite{Konoplya:2011qq,Berti:2009kk,Kokkotas:1999bd,Nollert:1999ji} $\omega_{0}$) and the blueshift amplification inside  (represented by the surface gravity at the Cauchy horizon $\kappa_i$) becomes important  \cite{Bony,Dyatlov:2011jd,Hintz:2016gwb,Hintz:2016jak,Cardoso:2017soq,Dias:2018ynt}.
While for asymptotically flat case the power-law decay of perturbations at late times in the exterior region is weak enough to balance the blueshift under the event horizon, asymptotically de Sitter case is qualitatively different. As was shown in \cite{Bony,Dyatlov:2011jd,Hintz:2016gwb,Hintz:2016jak} for some initial perturbation $\Phi_{0}$, the evolution of the scalar-field perturbations around Schwarzschild-de Sitter black hole and its Kerr-Newman generalization is bound by the relation:
\begin{equation}
|\Phi -\Phi_{0}| \leq C e^{- \alpha t},
\end{equation}
where $\alpha$ is the imaginary part of the least damped quasinormal mode. This is in full concordance with the numerical time-domain evolution of perturbations \cite{Brady:1996za,Brady:1999wd,Molina:2003dc}. Further in \cite{Cardoso:2017soq} it was shown that for scalar-field perturbations of the Reissner-Nordström-de Sitter black hole the above balance leads to the statement that in the black-hole spectrum must always exist the quasinormal mode respecting the following upper limit:
\begin{equation}\label{bound}
-\im{\omega_{0}}\leq\frac{\kappa_i}{2}.
\end{equation}
This observation induced extensive further studies of the quasinormal modes of black holes and their relation to the Strong Cosmic Censorship \cite{Cardoso:2018nvb,Dias:2018etb,Mo:2018nnu,Destounis:2018qnb,Liu:2019lon,Luna:2019olw,Ge:2018vjq,Destounis:2019omd,Carballo-Rubio:2021bpr,Gan:2019jac,Xu:2019wak,Casals:2020uxa,Churilova:2021nnc}.

On the one hand there exists a conjecture of S. Hod claiming that all physically realistic (dynamically
formed) black-hole spacetimes are characterized by the following upper bound \cite{Hod:2006jw}:
\begin{equation}\label{HodC}
-\im{\omega_{0}}\leq\frac{\kappa_0}{2},
\end{equation}
where $\kappa_0$ is the surface gravity of the event horizon. If one trusts this conjecture, it immediately follows that independently on the underlying gravitational theory \cite{Hod:2020ktb,Hod:2018lmi}
 \begin{equation}
-\im{\omega_{0}}\leq \frac{\kappa_0}{2}\leq\frac{\kappa_i}{2}.
\end{equation}
However, the Hod's conjecture (\ref{HodC}) itself requires a rigourous proof or even some numerical evidence and there are indications that it might be invalid for a number of cases (see, for example, \cite{Cuyubamba:2016cug,Churilova:2019sah}).

The current value of the cosmological constant $\Lambda = 1.0905 10^{-52} m^{-2}$ \cite{Planck:2018vyg} implies that the cosmological horizon is many orders larger than the radii of black holes, so that the regime of small values of $\Lambda$ is of the greatest interest.
Here we will present numerical evidence that in the regime of small values of the cosmological constant both the Hod's proposal (\ref{HodC}) and the strong cosmic censorship bound for quasinormal modes (\ref{bound}) are satisfied for general spherically symmetric black holes in an arbitrary metric theory of gravity. For this purpose we will analyze quasinormal spectra of general asymptotically de Sitter parametrized black holes.

\section{General parametrized spherically symmetric black hole in the de Sitter space}
Here we will extend the general parametrization of asymptotically flat spherically symmetric black holes in metric theories of gravity~\cite{Rezzolla:2014mua} to the asymptotically de Sitter case.
The metric of a spherically symmetric black hole can be written in the following general form,
\begin{equation}
ds^2=-N^2(r)dt^2+\frac{B^2(r)}{N^2(r)}dr^2+r^2 (d\theta^2+\sin^2\theta d\phi^2),\label{metric}
\end{equation}
where $r_0$ is the event horizon, so that $N(r_0)=0$.

Following \cite{Rezzolla:2014mua}, we will use the new dimensionless variable
$$x \equiv 1-\frac{r_0}{r},$$
so that $x=0$ corresponds to the event horizon, while $x=1$ corresponds to spatial infinity. We rewrite the metric function $N$ via the expression
$$N^2=x A(x),$$
where $A(x)>0$ for \mbox{$0\leq x\leq1$}. We represent the functions $A$ and $B$, as follows:
\begin{eqnarray}\label{Aexp}
A(x)&=&-\lambda(1-x)^{-2}-(\nu+\lambda)(1-x)^{-1}+\kappa
\\\nonumber&&-\epsilon (1-x)+(a_0-\epsilon)(1-x)^2+{\tilde A}(x)(1-x)^3\,,
\\
B(x)&=&1+b_0(1-x)+{\tilde B}(x)(1-x)^2\,.\label{Bexp}
\end{eqnarray}
where $\lambda$ and $\nu$ are the cosmological coefficients, corresponding to the cosmological constant,
$$\lambda=\frac{\Lambda r_0^2}{3}\geq0,$$
and the effective dark-matter term~$\nu$~\cite{Mannheim:1988dj}.
For the black hole of mass $60M_{\odot}$ one has $\lambda\approx10^{-42}$, while for the $SgrA^*$, this value is $\lambda\approx 5.6\times10^{-33}$.

We fix $\kappa\equiv1-\lambda-\nu$, and $\epsilon$ is related to the asymptotic mass
$$\epsilon=\frac{2M}{r_0}-\kappa=\frac{2M}{r_0}-1+\lambda+\nu,$$
while the coefficients $a_0$ and $b_0$ can be expressed in terms of the post-Newtonian parameters $\beta$ and $\gamma$,
$$a_0=(\beta-\gamma)\frac{(\kappa+\epsilon)^2}{2}, \qquad b_0=(\gamma-1)\frac{\kappa+\epsilon}{2}.$$
Current observational constraints on the PN parameters imply \mbox{$a_0 \sim b_0 \sim 10^{-4}$}, so that we can safely neglect them.

All the above coefficients describe the behaviour of the metric functions at large distance, particularly, $\lambda$ and $\nu$ are matched at the cosmological distances.
When $\lambda=\nu=0$ the function $A(x)$ in (\ref{Aexp}) coincides with parametrization of the spherically symmetric black hole in the asymptotically flat spacetime \cite{Rezzolla:2014mua}.

The functions ${\tilde A}$ and ${\tilde B}$ are introduced through the infinite continued fraction in order to describe the metric near the horizon (\ie for $x \simeq 0$),
\begin{equation}\label{ABdef}
{\tilde A}(x)=\frac{a_1}{1+\frac{a_2x}{1+\frac{a_3x}{1+\ldots}}}, \qquad
{\tilde B}(x)=\frac{b_1}{1+\frac{b_2x}{1+\frac{b_3x}{1+\ldots}}},
\end{equation}
where $a_1, a_2,\ldots$ and $b_1, b_2,\ldots$ are dimensionless constants to be constrained from observations of phenomena which are localized near the event horizon. At the horizon only the first term in each of the continued fractions survives,
$
{\tilde A}(0)={a_1},~
{\tilde B}(0)={b_1},
$
which implies that near the horizon only the lower-order terms of the expansions are essential.

The cosmological parameter $\lambda$ can be expressed in terms of the cosmological horizon $r_c>r_0$, such that $N(r_c)=0$.
When all the coefficients $a_i$, $b_i$, $\epsilon$, and $\nu$ vanish, we have the Schwarzschild-de Sitter black hole.

\section{Upper bound for the damping rate of the fundamental mode}
Quasinormal modes of the empty de Sitter spacetime can be found in analytic form~\cite{Lopez-Ortega:2006tjo,Lopez-Ortega:2006aal},
\begin{equation}\label{dSmodes}
\omega_{n}^{(dS)}r_{c} = -\imo (\ell + n+1-\delta_{s0}\delta_{n0}),
\end{equation}
where $s$ is the spin of the field,  $n=0,1,2,\ldots$ is the overtone number and $\ell=s,s+1,\ldots$ is the multiple number. For massless scalar ($s=0$) and electromagnetic ($s=1$) fields propagating in the background of the general spherically symmetric black hole we see that varying the parameters $\epsilon$, $a_1$, $b_1$ and others always respects the pure de Sitter limit of non-oscillatory quasinormal modes. In other words, when $r_{0}/r_{c} \to 0$, then the purely imaginary quasinormal frequencies go over into the above modes of the empty de Sitter spacetime.

Notice that, although $s=\ell=0$ case is not considered in \cite{Lopez-Ortega:2006tjo}, the dominant mode given by the formula~(\ref{dSmodes}), $\omega=0$ exists not only in the empty de Sitter space, but also in the Schwarzschild-de Sitter spectrum. In the time domain this nondynamic mode manifests itself as the settling down of the scalar-field profile to a constant value at late times~\cite{Brady:1999wd}. It seems that the mode $\omega=0$ exists for $s=\ell=0$ in the spectrum of the deformed asymptotically de Sitter black holes as well.

\begin{figure*}
\resizebox{\linewidth}{!}{\includegraphics{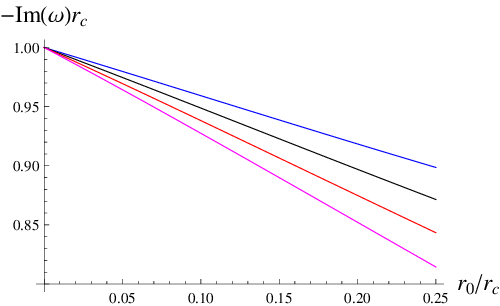}\includegraphics{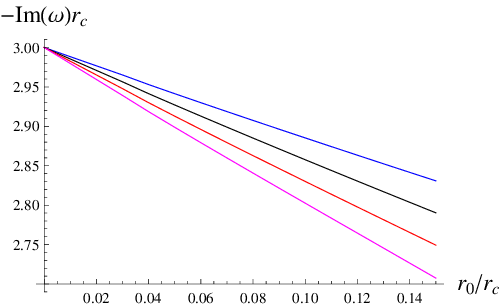}\includegraphics{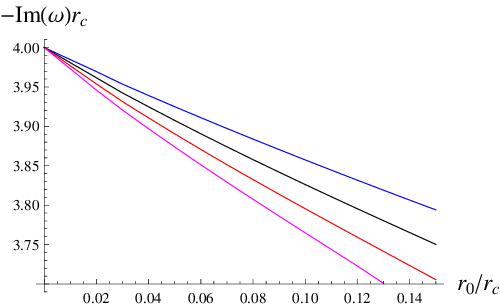}}
\caption{The dominant (pure imaginary) frequencies of the scalar field ($\ell=1$) as functions of $r_0/r_c$ for $a_0=a_1=0$, $b_0=b_1=0$, $\nu=0$, and $\epsilon=-0.2$ (blue), $\epsilon=0$ -- Schwarzschild-de Sitter (black), $\epsilon=0.2$ (red), $\epsilon=0.4$ (magenta).}\label{fig:epsilon}
\end{figure*}

\begin{figure}
\resizebox{\linewidth}{!}{\includegraphics{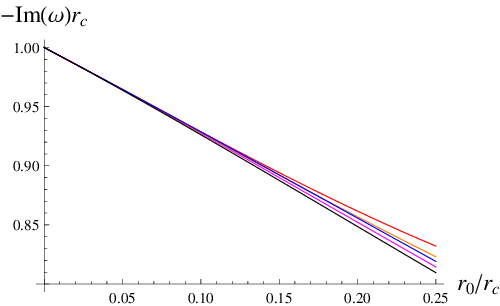}}
\caption{The dominant (pure imaginary) frequency of the scalar field ($\ell=1$) as functions of $r_0/r_c$ for $\nu=0$, $a_0=b_0=0$, and $\epsilon=0.4$, from bottom to top: $a_1=0$, $b_1=-0.5$, and $b_2=0$ (black), $a_1=b_1=0$ (magenta), $a_1=0$, $b_1=0.5$, and $b_2=0$ (blue), $b_1=0$, $a_1=1.0$, and $a_2=0$ (orange), $b_1=0$, $a_1=2.0$, and $a_2=0$ (red).}\label{fig:ab}
\end{figure}

Here we calculate quasinormal modes of black holes with the help of the Bernstein spectral method~\cite{Fortuna:2020obg}, which provides quite a good accuracy for purely imaginary frequencies \cite{Jansen:2017oag,Konoplya:2022xid}.
After calculation of quasinormal frequencies for various values of the deformation parameters we can summarize that for $r_{c} \gg r_{0}$ the dominant modes of the deformed Schwarzschild-de Sitter black holes, $\omega_n$ approach de Sitter modes~(\ref{dSmodes}) for any choice of the deformation parameters. The quasinormal modes in this regime obey a simple and universal law (see for example, Fig.~\ref{fig:epsilon}):
\begin{eqnarray}\label{BHmodes}
  \omega_n&=&\omega_{n}^{(dS)}\left(1-\frac{r_0(1+\epsilon)}{2r_c}+\Order{\frac{r_e}{r_c}}^2\right)
  \PRDstyle{\\\nonumber&=&}\JCAPstyle{=}
  \omega_n^{(dS)}\left(1-\frac{M}{r_c}+\Order{\frac{M}{r_c}}^2\right).
\end{eqnarray}

We have calculated a great number of various cases with non-zero and large values of the coefficients of the parametrization $\epsilon,\nu,a_0,a_1,a_2,\ldots,b_0,b_1,b_2,\ldots,$ and found that in the dominant order of $1/r_c$, the frequencies depend only on the asymptotic parameters, $\epsilon$, $\lambda$ and $\nu$, and do not depend on the post-Newtonian parameters $a_0$ and $b_0$ or the parameters $a_1,a_2,\ldots$, $b_1,b_2,\ldots$. As an illustration of the above claim, Fig.~\ref{fig:ab} shows that for the fixed value of $\epsilon$, the dominant mode approaches the de Sitter value according to the formula (\ref{BHmodes}) and does not depend on the near-horizon parameters $a_1$ and $b_1$ when $r_0\ll r_c$.

The spectrum (\ref{dSmodes}) is different for the cosmological models with $\nu \neq0$. Although such an analysis is beyond the scope of this letter, we expect that, when the term $\nu$ is non-zero, the dominant correction to the quasinormal modes depends again on the black-hole mass only and does not depend on the parameters $a_0,a_1,a_2,\ldots$ and $b_0,b_1,b_2,\ldots$. Time-domain evolution of perturbations in the presence of such effective dark matter term $\nu\neq0$ \cite{Konoplya:2020fwg} indicates that the non-oscillatory long-lived modes should exist in this case as well, supporting the above assumption.

This law is certainly broken for a black hole comparable with the cosmological horizon scale, that is, the one filling almost the whole Universe. Then the oscillating modes become dominant and the purely imaginary modes depend stronger on the near-horizon parameters.

Once $r_c \gg r_{0}$, there is always a non-oscillatory frequency which is close to $\omega_{0}r_{c} \approx -\imo\ell$ for the scalar and $\omega_{0}r_{c} \approx -\imo(\ell+1)$ for the electromagnetic and gravitational perturbations. Thus, for the lowest multipole we have
\begin{equation}
|Im (\omega_0)| \propto \frac{1}{r_{c}}\ll \frac{\kappa_0}{2}, \quad  r_{c} \gg r_{0},
\end{equation}
where we took into account that the surface gravity is
\begin{equation}
  \kappa_0=\frac{1}{2B(r_0)}\frac{dN^2}{dr}\Bigg|_{r=r_0}\propto\frac{1}{r_0}.
\end{equation}
Taking into account that the surface gravity at the Cauchy horizon is either larger or equal to that at the event horizon, we conclude that
\begin{equation}
|Im (\omega_0)| \ll \frac{\kappa_0}{2} \leq \frac{\kappa_i}{2}.
\end{equation}
Thus, we see that once $r_0 \ll r_{c}$ both the Strong Cosmic Censorship bound for quasinormal modes (\ref{bound}) and the Hod's proposal (\ref{HodC}) are guaranteed.

Here we considered only test fields, because gravitational perturbations cannot be determined without resorting to a particular gravitational theory. Nevertheless, it is evident that the wave equation of any gravitational perturbations in the limit of small (in comparison with the de Sitter radius) black holes will tend to the one for the empty de Sitter space (eq. \ref{dSmodes} at $s=2$) and, consequently, the frequencies will tend to non-oscillatory modes of the de Sitter space (\ref{dSmodes}). Thus, the above bounds (\ref{bound}) and  (\ref{HodC}) for test fields will be valid for gravitational perturbations as well. Nevertheless, the subdominant term, proportional to $M/r_{c}$, of the universal law for quasinormal frequencies (given by eq. \ref{BHmodes}) might, in general, be different.

\section{Conclusions}
Using the general spherically symmetric parametrization of black-hole spacetimes we provided numerical evidence that once a natural requirement that the event horizon is much smaller than the cosmological one is fulfilled, then the Strong Cosmic Censorship bound for quasinormal modes, suggested initially in the Einstein theory, must be satisfied in arbitrary metric theories of gravity. This versatility is expressed in a universal behavior of the low-lying quasinormal modes which depend only upon the asymptotically defined parameters of the spacetime, the value of the cosmological constant and the deviation of the black-hole radius from its Schwarzschild limit, and not on the near-horizon geometry. We have also shown that even stronger Hod's bound is guaranteed in the above regime.

An important question which was beyond our consideration is whether these bounds are true for an arbitrary (stationary) rotating and asymptotically flat black hole. For this purpose one could explore the general parametrization for axially symmetric black holes developed in \cite{Konoplya:2016jvv}, but it is almost definite that the same bound must be valid in the rotating case as well, because the limit of the empty de Sitter spectrum must be achieved when the radius of the black hole is much smaller than the de Sitter horizon. The same logic concerns gravitational perturbations \cite{Konoplya:2022xid} which we have not considered here, because for that the gravitational theory must be defined in each case of deformation: Since the gravitational perturbations must be reduced to the ones for pure de Sitter spacetime in the regime of small black holes, the bound is guaranteed for gravitational perturbations as well.

\acknowledgments
A.~Z. was supported by Conselho Nacional de Desenvolvimento Científico e Tecnológico (CNPq).
R.~K. thanks support of Czech Science Foundation (GAČR) grant 19-03950S.

\end{document}